\documentclass[epj,final]{svjour}
\usepackage{cite}
\usepackage{psfig}
\usepackage{epsfig}
\newcommand{\bbox}[1]{\mbox{\boldmath $#1$}}

\newcommand{\mean}[1]{\left\langle #1 \right\rangle}
\newcommand{\abs}[1]{\left| #1 \right|}

\renewcommand{\phi}{\varphi}
\renewcommand{\rho}{\varrho}

\newcommand{\non}{\nonumber \\}

\newcommand{\eqn}[1]{Eq. (\ref{#1})}
\newcommand{\eqs}[2]{Eqs. (\ref{#1}), (\ref{#2})}
\newcommand{\pic}[1]{Fig. \ref{#1}}

\newcommand{\name}[1]{{\rm #1}}
\newcommand{\D}{\displaystyle}

\begin{document}

\title{Brownian Particles far from Equilibrium}

\dedication{Dedicated to the 70th birthday of Youri M. Romanovsky}

\author{Udo Erdmann \and Werner Ebeling \and Lutz Schimansky-Geier \and
  Frank Schweitzer} \authorrunning{Erdmann et. al}

\institute{Institute of Physics, Humboldt University, Invalidenstra{\ss}e
  110, 10115 Berlin, Germany}

\mail{udo.erdmann@physik.hu-berlin.de}

\date{Submitted: November 23, 1999} 

\abstract{We study a model of Brownian particles which are pumped with energy
  by means of a non-linear friction function, for which different types are
  discussed. A suitable expression for a non-linear, velocity-dependent
  friction function is derived by considering an internal energy depot of the
  Brownian particles. In this case, the friction function describes the
  pumping of energy in the range of small velocities, while in the range of
  large velocities the known limit of dissipative friction is reached. In
  order to investigate the influence of additional energy supply, we discuss
  the velocity distribution function for different cases. Analytical solutions
  of the corresponding Fokker-Planck equation in 2d are presented and compared
  with computer simulations.  Different to the case of passive Brownian
  motion, we find several new features of the dynamics, such as the formation
  of limit cycles in the four-dimensional phase-space, a large mean squared
  displacement which increases quadratically with the energy supply, or
  non-equilibrium velocity distributions with crater-like form. Further, we
  point to some generalizations and possible applications of the model.}


\maketitle


\section{Introduction} 

As we know from classical physics, Brownian motion denotes the erratic motion
of a small, but larger than molecular, particle in a surrounding medium, e.g.
a gas or a liquid.  This erratic motion results from the random impacts
between the atoms or molecules of the medium and the (Brownian) particle,
which cause changes in the direction and the amount of its velocity,
$\bbox{v}$.

This type of motion would be rather considered as \emph{passive motion},
simply because the Brownian particle does not play an active part in this
motion. The Brownian particle keeps moving because the dissipation of energy
caused by friction is compensated by the stochastic force, as expressed in the
fluctuation-dissipation theorem (Einstein relation).

In this paper, we are particularly interested in the non-equilibrium motion of
Brownian particles. The out-of-equilibrium state is reached by considering an
additional influx of energy which is transfered into kinetic energy of the
particles.  This will be considered by a more complex friction coefficient,
$\gamma_{0}$, which now can be a space- and/or velocity-dependent function,
$\gamma(\bbox{r,v})$, which can be also negative.

\emph{Negative friction} is known i.e. from technical constructions, where
moving parts cause a loss of energy, which is from time to time compensated by
the pumping of mechanical energy. For example, in a mechanical clock the
dissipation of energy by moving parts is compensated by a heavy weight. The
weight stores potential energy, which is gradually transferred to the clock
via friction, i.e. the strip with the weight is pulled down. Another example
are violin strings, where friction-pumped oscillations occur, if the violin
bow transfers energy to the string via friction. Already in the classical work
``The Theory of Sound'' of Lord Rayleigh considered motion with energy supply
\cite{Ra45}.

Provided a supercritical influx of energy, passive motion could be transformed
into \emph{active motion}, which relies on the supply of energy from the
surroundings. Active motion is of interest for the dynamics of \emph{driven
  systems}, such as physico-chemical
\cite{VaRoChYa87,DeDu71,DuDe74,Ge85,MiMe97} or biological \cite{SchiGr93}
systems. However, recent models on self-driven particles often neglect the
energetic aspects of active motion while focusing on the interaction of
particles \cite{ViCzBeCoSh95,DeVi95}.
  
In order to describe active motion in the presence of stochastic forces, we
have suggested a model of \emph{active Brownian particles}, which have the
ability to take up energy from the environment, to store it in an internal
energy depot and to convert internal into kinetic energy
\cite{SchwEbTi98,EbSchwTi99,SchwTiEb99,TiSchwEb99,EbErSchiSchw99}. Two of us
have shown that these particles could move in an ``high-velocity'' or active
mode, which results in different types of complex motion.

Other previous versions of active Brownian particle models consider also
specific activities, such as environmental changes or signal-response
behavior. In particular, the active Brownian particles are assumed to generate
a self-consistent field which effects their further movement
\cite{SchwSchi94,SchiMiRoMa95,Er99}. The non-linear feedback between the
particles and the field generated by themselves may result in an interactive
structure formation process on the macroscopic level. Hence, these models have
been used to simulate a broad variety of pattern formations in complex
systems, ranging from physical to biological and social systems
\cite{La95,SchwLaFa97,HeSchwKeMo97,MiZa99}.

The main objective of this work is to study the influence of different types
of non-linear friction terms on active Brownian particles and the
corresponding probability distributions. Hence, we neglect possible
environmental changes of the particles and rather focus on the energetic
aspects of motion in the presence of noise. In Sect. \ref{sec:2}, we introduce
the idea of pumping by active friction and outline the basic dynamics of our
model by means of Langevin and Fokker-Planck equations.  Further, we
investigate the distribution function for spatially uniform systems, i.e.  in
the absence of an external potential. In Sect.  \ref{sec:3} we consider active
Brownian motion in external potentials. For the case of a parabolic potential,
we derive analytical expressions for the distribution function and present
computer simulations based on the Langevin equation, which show the occurrence
of limit cycles. The basic ideas are then generalized for other types of
potentials. Finally, we discuss some possible applications for the dynamics of
pumped Brownian particles.


\section{Active Brownian Motion of Free Particles}
\label{sec:2}
\subsection{Equations of Motion for Active Brownian Motion}
\label{ssec:2.1}
The motion of a Brownian particle with mass $m$, position $\bbox{r}$, and
velocity $\bbox{v}$ moving in a space-dependent potential, $U(\bbox{r})$, can
be described by the following Langevin equation:
\begin{equation}
  \label{langev-or}
  \dot{\bbox{r}}=\bbox{v}\,;\,\, 
  \dot{\bbox{v}}= -\gamma(\bbox{v})\, \bbox{v} 
  - \frac{1}{m} \bbox{\nabla} U(\bbox{r}) + {\cal F}(t)
\end{equation}
Here, $\gamma(\bbox{v})$ is the friction coefficient, which is assumed to
depend on velocity and and therefore implicitly on time. This rather general
ansatz will allow us to consider more complex functions for the friction
coefficient, as discussed below.  ${\cal F}(t)$ is a stochastic force with
strength $D$ and a $\delta$-correlated time dependence
\begin{equation}
  \label{stoch}
  \mean{{\cal F}(t)}=0 \,;\,\,
  \mean{{\cal F}(t){\cal F}(t')}=2D \,\delta(t-t')
\end{equation}
In the case of thermal equilibrium systems with
$\gamma(\bbox{v})=\gamma_{0}={\rm const.}$ we may assume that the loss of
energy resulting from friction, and the gain of energy resulting from the
stochastic force, are compensated in the average. In this case the
fluctuation-dissipation theorem (Einstein relation) applies:
\begin{equation}
  \label{fluct-diss}
  D=\gamma_{0} \frac{k_{B}T}{m}
\end{equation}
$T$ is the temperature and $k_{B}$ is the Boltzmann constant.

In this paper we are mainly interested in the discussion of the probability
density $P(\bbox{r},\bbox{v},t)$ to find the particle at location $\bbox{r}$
with velocity $\bbox{v}$ at time $t$. As well known, the distribution function
$P(\bbox{r},\bbox{v},t)$ which corresponds to the Langevin \eqn{langev-or},
can be described by a Fokker-Planck equation of the form:
\begin{eqnarray}
  \label{fpe-or}
  \frac{\partial P(\bbox{r},\bbox{v},t)}{\partial t} & =&
  \frac{\partial}{\partial \bbox{v}}
  \left\{\gamma(\bbox{v})\,\bbox{v}\, P(\bbox{r},\bbox{v},t) + D\,
    \frac{\partial P(\bbox{r},\bbox{v},t)}{\partial \bbox{v}}\right\} \non &&
  - \bbox{v}\, \frac{\partial P(\bbox{r},\bbox{v},t)}{\partial \bbox{r}} +
  \frac{1}{m} \nabla 
  U(\bbox{r}) \,\frac{\partial P(\bbox{r},\bbox{v},t)}{\partial\bbox{v}}
\end{eqnarray}
In the special case $\gamma(\bbox{v})=\gamma_{0}$ the stationary solution of
\eqn{fpe-or}, $P^{0}(\bbox{r},\bbox{v})$, is known to be the Boltzmann
distribution:
\begin{equation}
  \label{eq:maxw}
  P^{0}(\bbox{r},\bbox{v}) = 
  C \, \exp{ \left( -\frac{U(\bbox{r})}{k_{B}T}\right)} \;
  \exp{\left(- \frac{\gamma_{0}}{2\,D}\,\bbox{v}^2\right)}
\end{equation}
where the constant $C$ results from the normalization condition.  

The major question discussed throughout this paper is how this known picture
changes if we add a new degree of freedom to the model by considering that
Brown\-ian particles can be also pumped with energy from the environment. This
extension leads to the model of active Brownian particles
\cite{SchwSchi94,SchiMiRoMa95,EbSchwTi99,SchwEbTi98}. While the usual dynamic
approach to Brownian motion is based on the assumption of a \emph{passive
  friction} described by a constant, non-negative friction coefficient,
$\gamma_{0}>0$, active Brownian motion considers \emph{active friction} as an
additional mechanism of the particle to gain energy. In the present model,
this will be described by a more complex friction coefficient
$\gamma(\bbox{v})$ which is a velocity-dependent function. Several aspects of
an additional \emph{space} dependence were discussed in an earlier work
\cite{SteEbCa94}.

In the next section, we will introduce two possible ansatzes for the pumping
of energy, where the active friction depends on the velocity $\bbox{v}$ of the
particle.  In Sect. \ref{sec:2} we will restrict the discussion to the case
$U(\bbox{r})={\rm const.}$, but in Sect.  \ref{sec:3} also the influence of
external forces will be discussed.  Further, from now on we will choose units
in which $m \equiv 1$.

\subsection{Pumping By Velocity-Dependent Friction} 
\label{ssec:2.2}
The velocity-dependent pumping of energy plays an important role e.g. in
models of the theory of sound developed by \name{Rayleigh} \cite{Ra45} already
at the end of the last century.  In the simplest case we may assume the
following friction function for the Brownian particle:
\begin{equation}
  \label{gamma-ray}
  \gamma(\bbox{v}) = - \gamma_1 + \gamma_2\, \bbox{v}^2 
  =\gamma_1 \left(\frac{\bbox{v^2}}{\bbox{v}_0^2} -1\right)
\end{equation}
This Rayleigh-type model is a standard model for self-sustained oscillations
studied in different papers on nonlinear dynamics \cite{Kl94,MiZa99}. We note
that
\begin{equation}
  \label{v0-r}
  \bbox{v}_0^2 = \frac{\gamma_1}{\gamma_2}  
\end{equation}
defines a special value of the velocity where the friction function,
\eqn{gamma-ray}, is zero (cf.  \pic{gamma-ray-grul}).

\begin{figure}[htbp]
  \centerline{\psfig{figure=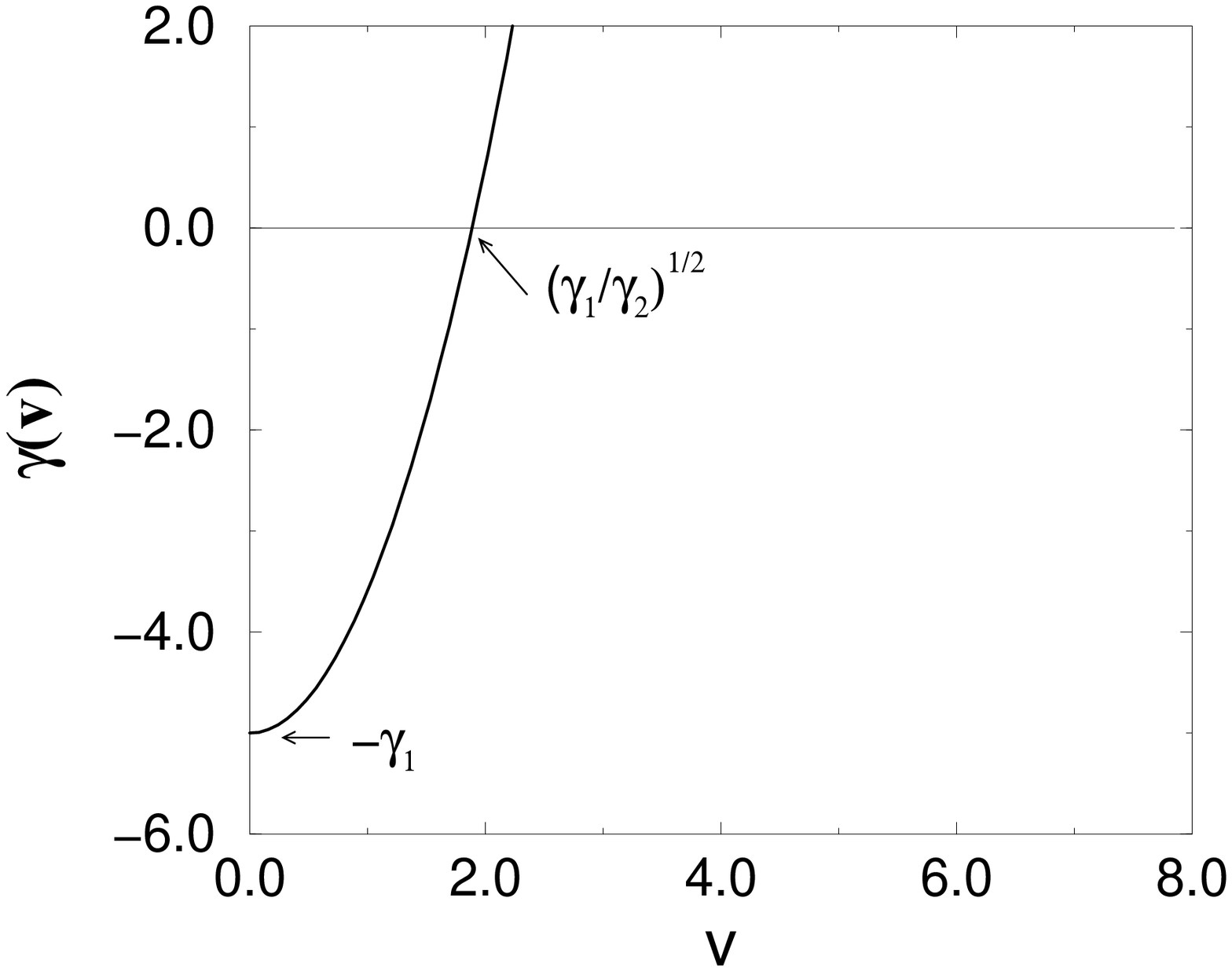,width=7cm}}
  \centerline{\psfig{figure=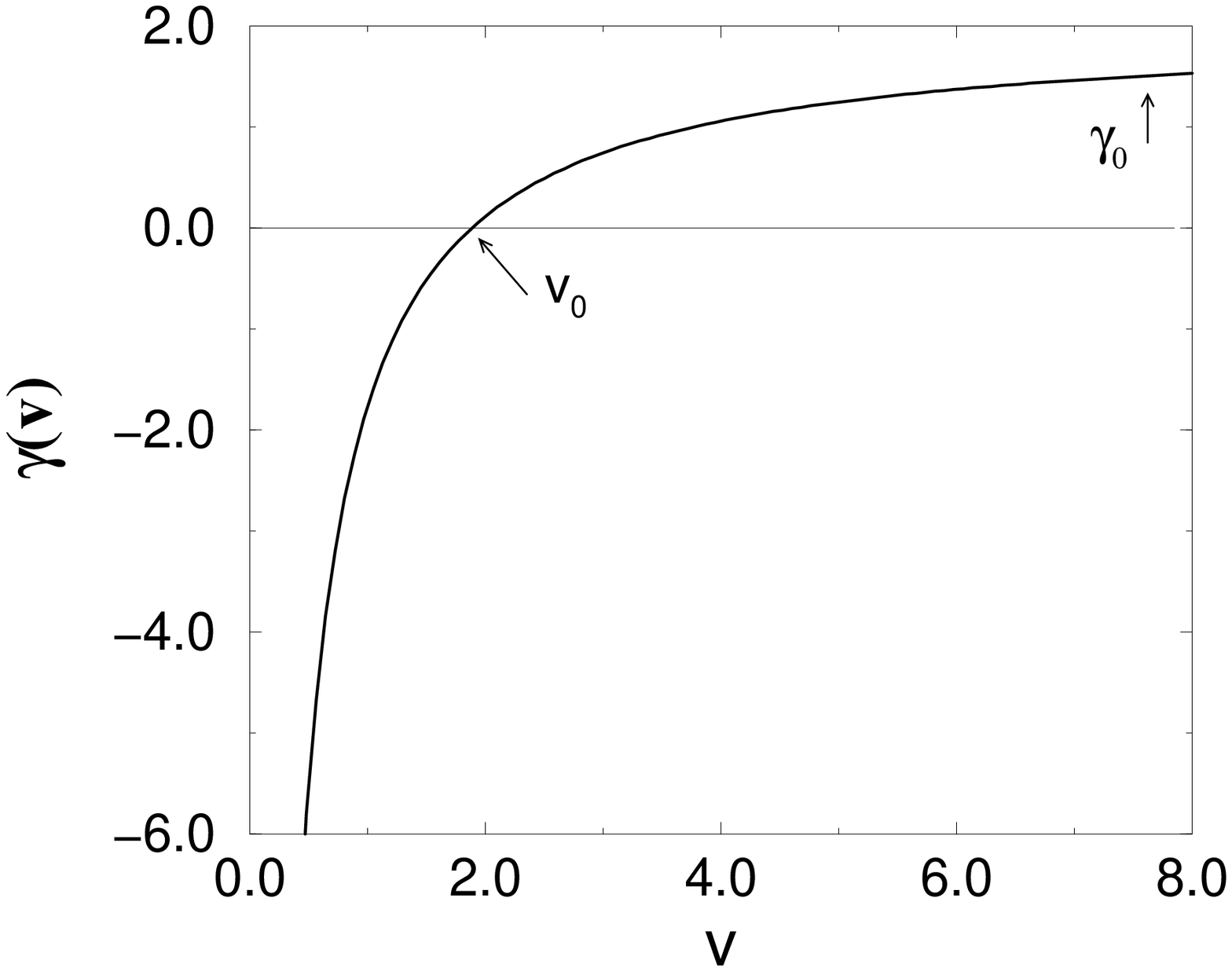,width=7cm}}
  \caption[]{\label{gamma-ray-grul}
    Two different types of velocity dependent friction functions: (top)
    Rayleigh type, \eqn{gamma-ray}, (bottom) \eqn{gamma-grul}.
    Parameters: $\gamma_{1}=5$, $\gamma_{2}=1.4$, $\gamma_{0}=2$,
    $v_{0}=1.889$. }
\end{figure}

Another standard model for active Brownian dynamics with a zero $\bbox{v}_0$
in the friction function introduced in \cite{SchiGr93} reads (cf.
\pic{gamma-ray-grul}):
\begin{equation}
\label{gamma-grul}
  \gamma(\bbox{v}) = \gamma_0 \left(1 - \frac{v_0}{v} \right)
\end{equation}
It has been shown that \eqn{gamma-grul} allows to describe the active motion
of different cell types, such as granulocytes \cite{SchiGr93,FrGr90,GrBu84}
monocytes \cite{BoRaGr89} or neural crest cells \cite{GrNu91}. Here, the speed
$v_{0}$ expresses the fact that the motion of cells is not only driven by
stochastic forces, instead cells are also capable of active motion.

To compare \eqs{gamma-ray}{gamma-grul} (cf. \pic{gamma-ray-grul}), we note
that in both ansatzes we have a certain range of (small) velocities, where the
friction coefficient can be negative, i.e. the motion of the particle can be
pumped with energy.  Due to the pumping mechanism, the conservation of energy
clearly does not hold for the particle, i.e. we now have a strongly
non-equilibrium canonical system \cite{FeEb89}.

Above a critical value of the velocity, described by $\bbox{v}_{0}$, the
active friction turns into passive friction again. In \eqn{gamma-ray}, the
increase of the friction with increasing velocity is not bound to a maximum
value.  However, this is the case in \eqn{gamma-grul}, where the friction
function converges into the constant of passive friction, $\gamma_{0}$, for
large $\bbox{v}$. The disadvantage of \eqn{gamma-grul}, on the other hand, is
the singularity of the friction function for $\bbox{v} \to 0$. In Sect.
\ref{ssec:2.3}, we will suggest a model of active friction which avoids these
drawbacks, while providing a more general approach to derive a suitable
function $\gamma(\bbox{v})$.

But before, we want to discuss some features which result from the Rayleigh
ansatz, \eqn{gamma-ray}. The stationary solution of the corresponding
Fokker-Planck \eqn{fpe-or} reads in the isotropic case:
\begin{equation}
  \label{RayHelm_stat}
  P^{0}(\bbox{v}) = C \exp\left(\frac{\gamma_1}{2D}
    \,\bbox{v}^2-\frac{\gamma_2}{4D} \, \bbox{v}^4\right)
\end{equation}
The distribution $P^{0}(\bbox{v})$, \eqn{RayHelm_stat}, has its maximum for
$\bbox{v}=0$, if $\gamma_{1} \leq 0$. For $\gamma_{1} > 0$, the mean of the
velocity distribution is still zero, but the maxima are different.
Considering a two-dimension motion, the distribution then has the form of a
hat (cf. also \pic{hat}), indicating that the particle most likely moves with
a constant absolute velocity $\bbox{v}^2=\bbox{v}_0^2=\gamma_{1}/\gamma_{2}$.

For the calculation of the normalization constant $C$ in \eqn{RayHelm_stat} we
use a method described by \cite{St67} and find with $\gamma_2 = 1$ the
explicit expression \cite{Er97}:
\begin{equation}
  C^{-1}=\pi \sqrt{\pi D}\; \exp \left( \frac{\gamma_1 ^2}{4D}%
  \right) \left[ 1+{\rm erf}\left( \frac {\gamma_1} {2\sqrt{D}}\right)
  \right]  
\label{normierung-stat-2d}
\end{equation}
Using \eqs{RayHelm_stat}{normierung-stat-2d}, we are able to calculate the
different moments of the stationary velocity distribution and find:
\begin{equation}
  \label{v-2n}
  \mean{v^{2n}} = \frac{(2D)^n}{Z_{2D}(\gamma_1 )}\;\frac{\partial^n}
  {\partial \gamma_1 ^n}\; Z_{2\mathrm{d}}(\gamma_1 )
\end{equation}
with the function:
\begin{equation}
 \label{z-2d}
  Z_{2\mathrm{d}}(\gamma_1) =\pi \sqrt{\pi D}\; \exp \left(
  \frac{\gamma_1 ^2}{4D }\right) \left[ 1+{\rm erf}\left( \frac
    {\gamma_1} {2\sqrt{D }}\right) \right]
\end{equation}
In particular, the second moment, which is proportional to the temperature,
and the fourth moment which is proportional to the fluctuations of the
temperature, read:
\begin{eqnarray}
 \mean{\bbox{v}^2} &=&\gamma_1 +\sqrt{\frac D \pi }\; \frac{2\exp \left( 
     \frac{\D \gamma_1 ^2}{\D 4D }\right) }{1+{\rm erf}\left( 
     \frac {\D \gamma_1} {\D 2\sqrt{D}}\right) } \non
 \left\langle \bbox{v}^4\right\rangle &=&\gamma_1 ^2+2D 
 + \gamma_1 \mean{\bbox{v}^2}
\end{eqnarray}
We note that a similar discussion can be carried out also for the case of the
friction function, \eqn{gamma-grul}. In this case the stationary solution of
the Fokker-Planck \eqn{fpe-or} is of particular simplicity \cite{SchiGr93}:
\begin{equation}
  \label{Gruler_stat}
  P^{0}(\bbox{v}) = C \exp\left(\frac{\gamma_0}{2D}\, 
    \left(v - v_0\right)^2\right)
\end{equation}
In the following section, we will derive a model which allows us to generalize
both special cases introduced here, so that the further discussion can be
unified.
 
\subsection{Pumping from an Internal Energy Depot}
\label{ssec:2.3}
In order to avoid the drawbacks of the friction functions,
\eqs{gamma-ray}{gamma-grul}, mentioned in the previous section, we need a
friction function $\gamma(\bbox{v})$ which does not diverge in the limit of
small or large velocities. Further the case of the ``normal'' passive friction
should be included as a limiting case.  The ansatz suggested here is based on
the recently developed model of Brownian motion with internal energy depot
\cite{SchwEbTi98,EbSchwTi99,TiSchwEb99,SchwTiEb99}.

We assume that the Brownian particle should be able to take up external
energy, which can be stored in an internal energy depot, $e(t)$.  This energy
depot may be may be altered by three different processes:
\begin{enumerate}
\item take-up of energy from the environment; where $q$ is the pump rate of
  energy,
\item internal dissipation, which is assumed to be proportional to the
  internal energy. Here the rate of energy loss, $c$, is assumed to be
  constant,
\item conversion of internal energy into motion, where
  $d\left(\bbox{v}^2\right)$ is the rate of conversion of internal to kinetic
  degrees of freedom and $\eta\left(\bbox{v}^2\right)$ is the efficiency of
  conversion. This means that the depot energy may be used to accelerate
  motion of the particle.
\end{enumerate}
This extension of the model is motivated by investigations of active
biological motion, which relies on the supply of energy. This energy can be
dissipated by metabolic processes, but can be also converted into kinetic
energy. The resulting balance equation for the internal energy depot of a
pumped Brownian particle is then given by:
\begin{equation}
  \frac{d}{dt} e(t) = q - c\;e(t) - \eta(\bbox{v}^2)d(\bbox{v}^2)\;e(t)
  \label{dep0}
\end{equation}
Generalizing the ansatz given in \cite{EbSchwTi99,SchwEbTi98} we postulate the
following Langevin equation for the motion of Brownian particles with internal
energy depot:
\begin{equation}
  \dot{\bbox{v}}+\gamma_0 \bbox{v} + \nabla U =
  \frac{\eta\left(\bbox{v}^2\right)d\left(\bbox{v}^2\right)}{\bbox{v}^2}\; 
  e(t)\bbox{v}+{\cal F}(t)
  \label{langeq-depot}
\end{equation}
Let us consider now two limiting cases:

\textbf{(i)} First, we assume
\begin{equation}
  q(\bbox{r})\equiv q_0 \; ;\quad d\left(\bbox{v}^2\right) = d_2 \bbox{v}^2
\; ;\quad \eta\left(\bbox{v}^2\right) \equiv 1
  \label{dv}
\end{equation}
where $d_2>0$ is the conversion factor. Using the assumption of a fast
relaxation of the internal energy depot we find the quasistationary value:
\begin{equation}
\label{e-0}
  e_{0}= \frac{q_0}{c+d_2 \bbox{v}^2}
\end{equation}
With the quasistationary approximation $e_{0}$, \eqn{e-0}, the internal energy
can be adiabatically eliminated, and we find the following effective friction
function:
\begin{equation}
 \label{gamma-v2}
  \gamma\left(\bbox{v}^2\right)= \gamma_0-\frac{q_0\,d_{2}}{c
+d_2 \bbox{v}^2}\;.
\end{equation}
which is plotted in \pic{gamma-pic}. We see that in the range of small
velocities pumping due to negative friction occurs, as an additional source of
energy for the Brownian particle. Hence, slow particles are accelerated, while
the motion of fast particles is damped.

\begin{figure}[htbp]
  \centerline{\psfig{figure=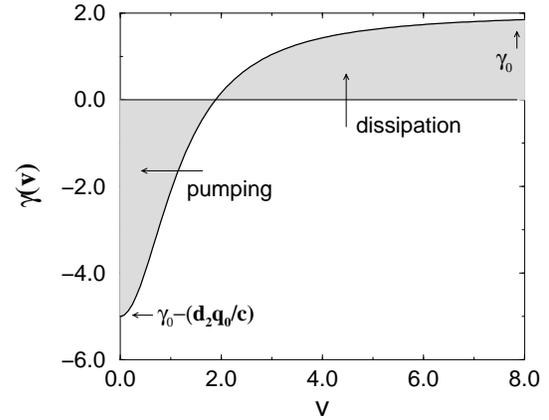,width=7.cm}}
  \caption[]{\label{gamma-pic}
    Velocity-dependent friction function $\gamma(v)$, \eqn{gamma-v2} vs.
    velocity $v$. The velocity ranges for ``pumping'' ($\gamma(v)<0$) and
    ``dissipation'' ($\gamma(v)>0$) are indicated.  Parameters: $q_0=10$;
    $c=1.0$; $\gamma_{0}=2$, $d_{2}=0.7$.}
\end{figure}

The friction function, \eqn{gamma-v2}, has again a zero, which reads in the
considered case:
\begin{equation}
  \label{v-0}
  \bbox{v}_0^2=\frac{q_0}{\gamma_0} - \frac{c}{d_2}
\end{equation}
\eqn{gamma-v2} agrees with the Rayleigh ansatz, \eqn{gamma-ray}, in the limit
of rather small velocities, $\bbox{v}\leq \bbox{v}_0$. From the conditions
$\gamma(\bbox{v}=0)$ and $\gamma(\bbox{v})=0$, we find for the two parameters
of the Rayleigh ansatz:
\begin{equation}
  \label{gam1-2}
  -\gamma_1=\gamma_0-q_{0}\frac{d_{2}}{c} \;\;; \quad 
  \gamma_2=\gamma_0 \frac{d_2}{c}
\end{equation}
Thus, the further discussion will also cover this case as long as $v\leq
v_{0}$.

\textbf{(ii)} In the second limiting case we consider a very large energy
depot which is then assumed to be constant, $e(t)=e_0={\rm const.}$ However,
we may assume that the efficiency of conversion,
$\eta\left(\bbox{v}^2\right)$, decreases with increasing velocity as follows:
\begin{equation}
  \eta\left(\bbox{v}^2\right)=\frac{\eta_1}{1+\eta_2 \bbox{v}^2}
\end{equation}
where $\eta_{1}$ and $\eta_{2}$ are constants.  With $\eta_{0}=\eta_{1}d_{2}$,
this case leads to the effective friction function \cite{TiSchwEb99}:
\begin{equation}
  \label{gamma-eta} 
  \gamma\left(\bbox{v}^2\right)=\gamma_0 - 
  \frac{\eta_0\;e_0}{1+\eta_2 \bbox{v}^2}
\end{equation}
Both expressions, \eqn{gamma-v2} and \eqn{gamma-eta}, can be brought into the
same form by defining some constants:
\begin{eqnarray}
  \label{gamma-equal}
  \gamma\left(\bbox{v}^2\right) &=& \gamma_0 \; \left[
    1- \frac{q^{\star}}{1+d^{\star}\bbox{v}^2} \right]  \\ 
  q^{\star}
  &=&\frac{q_{0}d_{2}}{c\gamma_{0}}=\frac{\eta_{0}e_{0}}{\gamma_{0}}\;;\; 
  d^{\star}=\frac{d_{2}}{c}=\eta_{2} 
\end{eqnarray}
Using further the expression for the stationary velocity $\bbox{v}_{0}^{2}$,
\eqn{v-0}, we find for \eqn{gamma-equal}
\begin{equation}
  \label{gamma-fin}
  \gamma\left(\bbox{v}^2\right) = \gamma_0 \; 
  \frac{\left(\bbox{v}^2-\bbox{v}_{0}^{2}\right)}{
    (q_{0}/\gamma_{0})+\left(\bbox{v}^2-\bbox{v}_{0}^{2}\right)}
\end{equation}
We restrict the further discussion to the two-dimensional space
$\bbox{r}=\{x_{1},x_{2}\}$. Then the stationary velocities $v_0$,
\eqs{v0-r}{v-0}, where the friction is just compensated by the energy supply,
define a cylinder in the four-dimensional space:
\begin{equation}
\label{cylinder}
  v_1^2 + v_2^2 = \bbox{v}_0^2
\end{equation}
which attracts all deterministic trajectories of the dynamic system. The
stationary solution of the Fokker-Planck \eqn{fpe-or} reads for the friction
function, \eqn{gamma-v2}, and in the absence of an external potential, i.e.
$U(x_{1},x_{2})\equiv 0$:
\begin{equation}
  \label{p0-v}
  P^{0}(\bbox{v})= C'\,\left(1+\frac{d_{2}\bbox{v}^2}{c}\right)
  ^{\frac{q_{0}}{2D}}\;
  \exp{\left( - \frac{\gamma_{0}}{2D} \,v^{2}\right)}
\end{equation}
Compared to \eqn{eq:maxw}, which describes the Maxwellian velocity
distribution of ``simple'' Brownian particles, a new prefactor appears now in
\eqn{p0-v} which results from the internal energy depot. In the range of small
values of $v^{2}$, the prefactor can be expressed by a power series truncated
after the first order, and \eqn{p0-v} reads then:
\begin{equation}
  \label{p0-v-app}
  P^{0}(\bbox{v}) \sim \exp{\left[ - \frac{\gamma_{0}}{2D} \left(
        1-\frac{q_{0}d_{2}}{c\gamma_{0}}\right)\,\bbox{v}^{2} + \cdots \right]}
\end{equation}
In \eqn{p0-v-app}, the sign of the expression in the exponent depends
significantly on the parameters which describe the balance of the energy
depot. For a subcritical pumping of energy, $q_{0}d_{2}< c \gamma_{0}$, the
expression in the exponent is negative and an \emph{unimodal velocity
  distribution} results, centered around the maximum $\bbox{v}_0=0$. This is
the case of the ``low velocity'' or \emph{passive mode} for the stationary
motion which here corresponds to the \emph{Maxwellian velocity distribution}.
However, for supercritical pumping,
\begin{equation}
  \label{critical}
  q_{0}d_{2}> c \gamma_{0}  
\end{equation}
the exponent in \eqn{p0-v-app} becomes positive, and a \emph{crater-like
  velocity distribution} results.  This is also shown in \pic{hat}. The maxima
of $P^{0}(\bbox{v})$ correspond to the solutions for $\bbox{v}_{0}^{2}$,
\eqn{v-0}.  The corresponding ``high velocity'' or \emph{active mode}
\cite{SchwTiEb99} for the stationary motion is described by strong deviations
from the Maxwell distribution.

\begin{figure} 
  \centerline{\psfig{figure=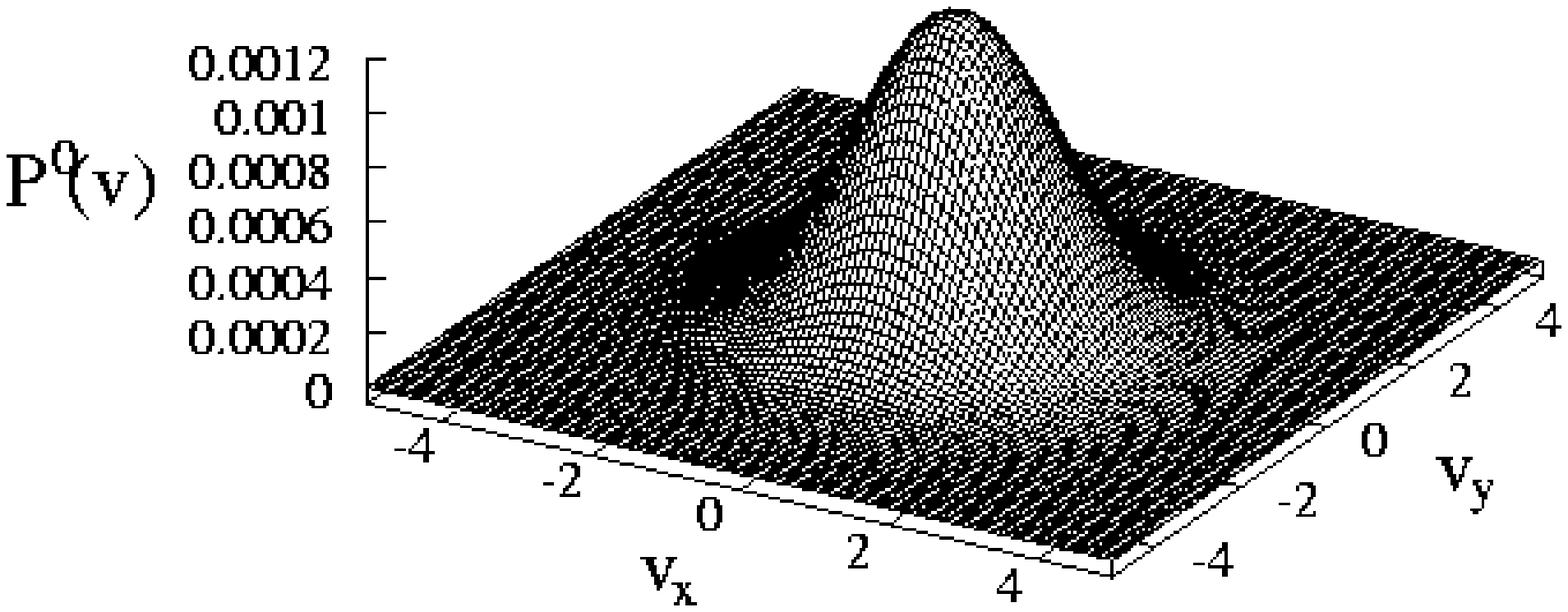,width=8cm}}
  \centerline{\psfig{figure=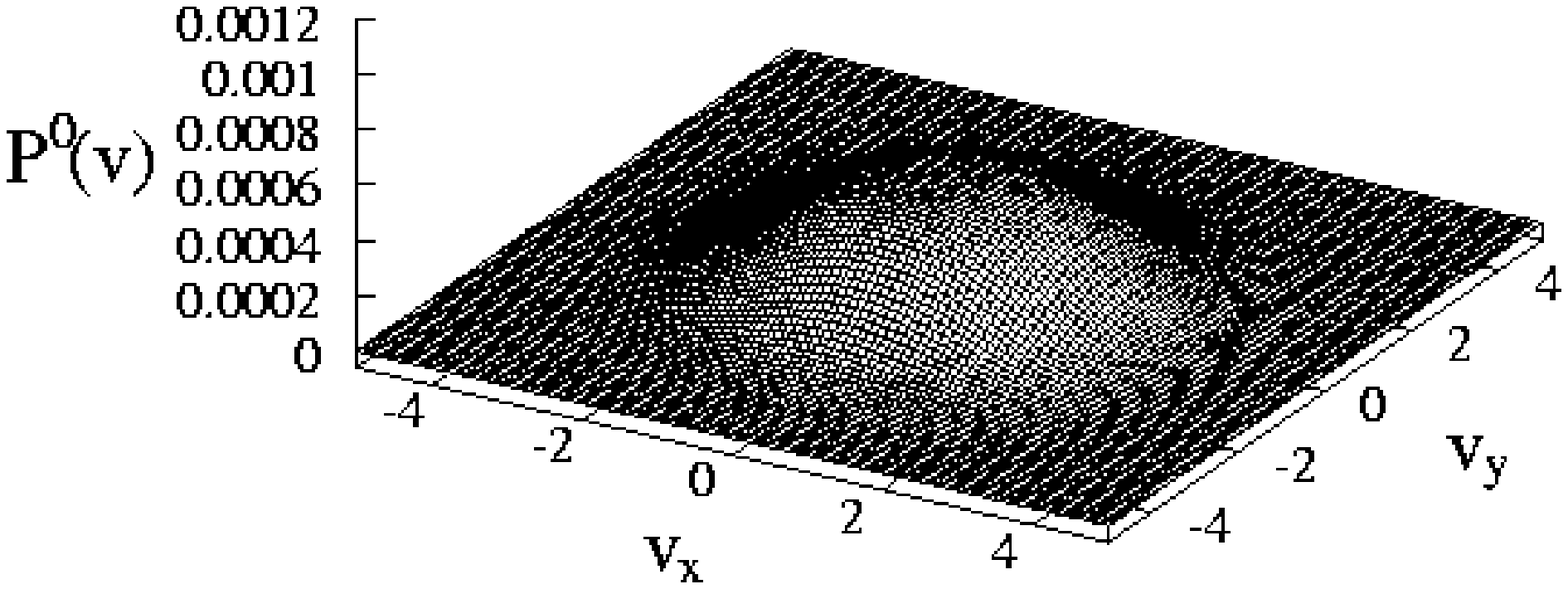,width=8cm}}
  \centerline{\psfig{figure=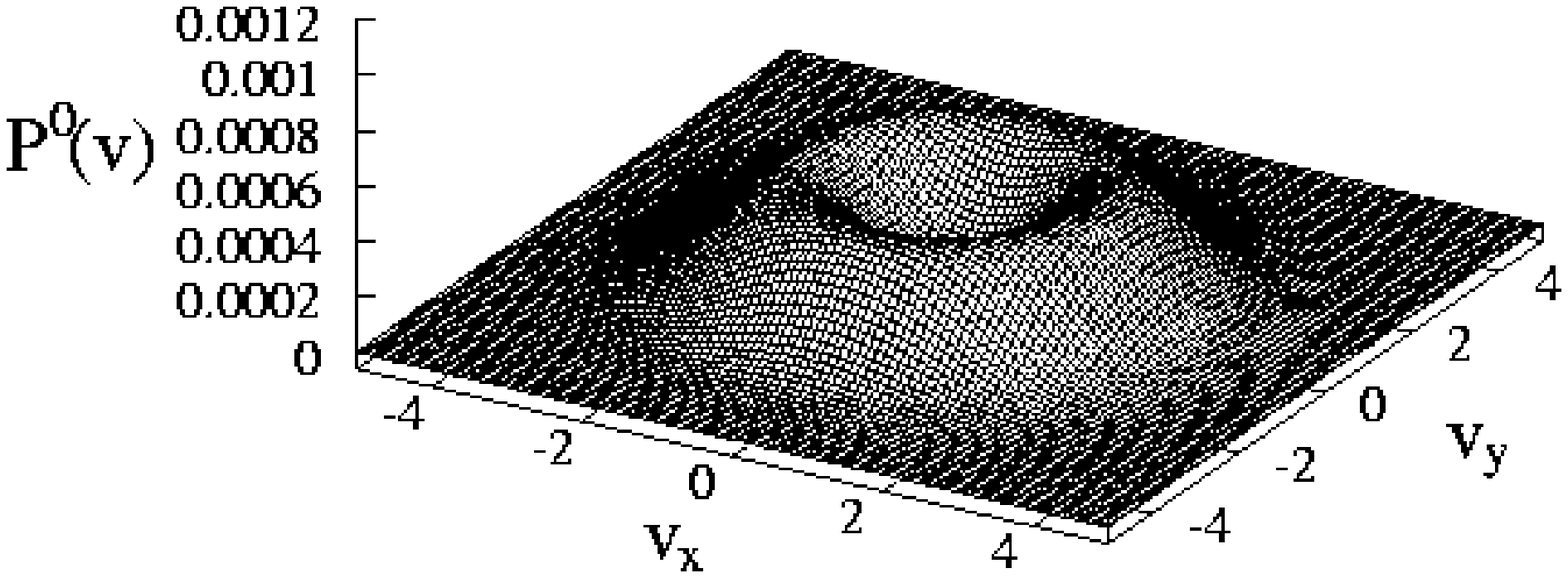,width=8cm}}
  \caption{\label{hat} 
    Normalized stationary solution $P^{0}(\bbox{v})$, \eqn{p0-v}, for $d_{2} =
    0.07$ (top), $d_{2} = 0.2$ (middle) and $d_{2}=0.7$ (bottom).  Other
    parameters: $\gamma_{0}=2$, $D=2$, $c=1$, $q_{0}=10$. Note that $d_{2} =
    0.2$ is the bifurcation point for the given set of parameters.}
\end{figure}

\subsection{Investigation of two Limit Cases}
\label{ssec:2.4}
In the limit of strong noise $D \sim T \rightarrow \infty$, i.e. at high
temperatures, we get from \eqn{p0-v} the known Maxwellian distribution by
means of \eqn{fluct-diss}:
\begin{equation}
  P^{0}(\bbox{v}) = \left(\frac{1}{2\pi k_B T}\right)
  \;\exp\left(-\frac{\bbox{v}^2}{2 k_B T}\right)
\end{equation}
which corresponds to standard Brownian motion in two dimensions. Hence, many
characteristic quantities are explicitly known as e.g. the dispersion of the
velocities
\begin{equation}
  \mean{\bbox{v}^2} = 2 k_B T
\end{equation}
and the Einstein relation for mean squared displacement
\begin{equation}
  \mean{\Big(\bbox{r}(t)-\bbox{r}(0)\Big)^2} = 
4 \frac{k_{B}T}{\gamma_{0}}\,t
\end{equation}
In the other limiting case of strong activation, i.e. relatively weak noise $D
\sim T \rightarrow 0$ and/or strong pumping, we find a $\delta$-distribution
instead:
\begin{equation}
  P^{0}(\bbox{v}) = C \delta\left(\bbox{v}^2 - \bbox{v}_0^2\right) \;;
  \quad \mean{\bbox{v}^2}=\bbox{v}_0^2
\end{equation}
In order to treat this case in the full phase space, we follow
\cite{SchiGr93,MiMe97} and introduce first an amplitude-phase representation
in the velocity-space:
\begin{equation}
  v_1 =v_0\,\cos{(\phi)} \;;\quad
  v_2 =v_0\,\sin{(\phi)}
\end{equation}
This allows us to separate the variables and we get a distribution function of
the form:
\begin{equation}
  P(x_1,x_2,v_1,v_2,t) = P(x_1,x_2,t) \cdot \delta(v_1^2 + v_2^2 - v_0^2) 
  \cdot  P\left( \phi ,t\right)
\end{equation}
The distribution of the phase $\phi$ satisfies the Fokker-Planck equation:
\begin{equation}
  \label{phidiffusion}
  \frac \partial {\partial t}P\left( \phi ,t\right) =D_{\phi}
  \frac{\partial ^2}{\partial \phi ^2}P\left( \phi ,t\right)
\end{equation}
By means of the known solution of \eqn{phidiffusion} we get for the mean
square:
\begin{equation}
   \label{phi-2}
   \mean{\phi ^2(t)} = D_{\phi}t   \;;\quad
   D_{\phi}=\frac {D}{v_0^2}  
\end{equation}
where $D_{\phi}$ is the angular diffusion constant.  By means of this, the
mean squared spatial displacement $\mean{r^2(t)}$ of the particle can be
calculated according to \cite{MiMe97} as:
\begin{eqnarray}
 \label{mean_r^2}
 \mean{r^2(t)} =\frac{2v_0^4t}D +\frac{v_0^6}{D^2 }\left[
   \exp \left( -\frac{2D t}{v_0^2}\right) -1\right]
\end{eqnarray}
For times $t\gg v_0^2/D$, we find from \eqn{mean_r^2} the following expression
for the effective spatial diffusion constant:
\begin{equation}
  \label{D_eff}
  D_r^{\rm eff} = \frac{2v_0^4}D = 
  \frac{2}{D}\left(\frac{q_{0}}{\gamma_{0}}-\frac{c}{d_{2}} \right)^{2}
\end{equation}
where $v_{0}$, \eqn{v-0}, considers the additional pumping of energy resulting
from the friction function, $\gamma(\bbox{v})$. Due to this additional
pumping, we obtain a high sensitivity with respect to noise expressed in the
scaling with $(1/D)$.


\section{Active Brownian Motion in External Potentials}
\label{sec:3}
\subsection{Motion in a Parabolic Potential}
In the following, we discuss the motion in a two-dimensional external
potential, $U(x_{1},x_{2})$, which results in additional forces on the pumped
Brownian particle. The case of a constant external force was experimentally
and theoretically investigated in \cite{SchiGr93}. Here we concentrate first
on a simple non-linear potential:
\begin{equation}
  U(x_1,x_2) = \frac{1}{2} a \,(x_1^2 + x_2^2)
  \label{parab}
\end{equation}
If we for the moment restrict the discussion to a deterministic motion, the
dynamics will be described by four coupled first-order differential equations:
\begin{equation}
  \begin{array}{rcl}
    \dot{x}_1 =  v_1 & \qquad &
    \dot{v}_1 =  - \gamma\left(v_1,v_2\right) v_{1} - ax_{1} \nonumber\\ 
    \dot{x}_2 =  v_2 & &  
    \dot{v}_2 =  - \gamma\left(v_1,v_2\right) v_2 - a x_{2}
  \end{array}
  \label{2d-det}
\end{equation}
For the one-dimensional Rayleigh model it is well known that this system
processes a limit cycle corresponding to sustained oscillations with the
energy $E_0=\gamma_1/\gamma_2$.  For the two-dimensional case we have shown
\cite{EbSchwTi99} that limit cycles occur in the four-dimensional space.  The
projection of this periodic motion to the $\{v_1,v_2\}$ plane is again the
cylinder, defined by \eqn{cylinder}. The projection to the $\{x_1,x_2\}$ plane
corresponds to a circle
\begin{equation}
  \label{x1-x2-const}
  x_1^2 + x_2^2 =  r_0^2 = {\rm const.}
\end{equation} 
The energy for motions on the limit cycle is: 
\begin{equation}
  \label{energy}
  E_0 = \frac{1}{2}(v_1^2 + v_2^2) +\frac{a}{2} (x_1^2 + x_2^2)
  = \frac{1}{2} v_0^2 +\frac{a}{2} r_0^2
\end{equation}
In \cite{EbSchwTi99} we have shown that any initial value of the energy
converges (at least in the limit of strong pumping) to
\begin{equation}
  H\longrightarrow E_0 =  v_0^2
\end{equation}
This corresponds to an equal distribution between kinetic and potential
energy. As for the harmonic oscillator in one dimension, both parts contribute
the same amount to the total energy. This result was obtained in
\cite{EbSchwTi99} using the assumption that the energy is a slow (adiabatic)
variable which allows a phase average with respect to the phases of the
rotation.

In explicite form we may represent the motion on the limit cycle in the
four-dimensional space by the four equations:
\begin{equation}
  \label{4d-explizit}
  \begin{array}{rcl}
    x_1 = r_0\, \cos(\omega t + \phi_0) & \qquad &
    v_1 = - r_0\, \omega\, \sin(\omega t + \phi_0)\\ \nonumber
    x_2 = r_0\, \sin(\omega t + \phi_0) &\qquad&
    v_2 = r_0\, \omega\, \cos(\omega t + \phi_0)
  \end{array}
\end{equation}
The frequency $\omega$ is given by the time the particle needs for one period
while moving on the circle with radius $r_0$ with constant speed $v_0$. This
leads to the relation
\begin{equation}
  \label{a-12}
  \omega_0 = \frac{v_0}{r_0} = \sqrt{a} = \omega
\end{equation}
This means that even in the case of strong pumping the particle oscillates
with the frequency given by the linear oscillator frequency $\omega$.

The trajectory defined by the four equations (\ref{4d-explizit}) is like a
hoop in the four-dimensional space. Therefore, most projections to the
two-dimensional subspaces are circles or ellipses. However there are two
subspaces, namely $\{x_1,v_2\}$ and $\{x_2,v_1\}$, where the projection is
like a rod.

A second limit cycle is obtained by time reversal:
\begin{equation}
  t \rightarrow -t; \quad v_1 \rightarrow -v_1 ; \quad v_2 \rightarrow -v_2
\end{equation}
This limit cycle also forms a hula hoop which is different from the first one
in that the projection to the $\{x_1,x_2\}$ plane has the opposite rotation
direction compared to the first one.  However both limit cycles have the same
projections to the $\{x_1,x_2\}$ and to the $\{v_1,v_2\}$ plane. The
separatrix between the two attractor regions is given by the following plane
in the four-dimensional space:
\begin{equation}
  \label{separat}
  v_1 + v_2 = 0
\end{equation}
Applying similar arguments to the \emph{stochastic} problem we find that the
two hoop-rings are converted into a distribution looking like two embracing
hoops with finite size, which for strong noise converts into two embracing
tires in the four-dimensional space \cite{EbErSchiSchw99}.  The projections of
the distribution to the $\{x_1,x_2,v_2\}$ subspace and to the
$\{v_1,v_2,x_2\}$ subspace are two-dimensional rings as shown in
\pic{fig:limit-cyc}. While in the deterministic case either left- or
righthanded rotations are found, in the stochastic case the system may switch
randomly between the left- and righthand rotations, since the separatrix
becomes penetrating (see \pic{fig:strongnoise}). This can be also confirmed by
looking at the projections of the distribution to the $\{x_1,v_2\}$ plane and
the $\{x_2,v_1\}$ plane (cf. \pic{fig:poincare}).  Since the hula hoop
distribution intersects perpendicular the $\{x_1,v_2\}$ and the $\{x_2,v_1\}$
plane, the projections to these planes are rod-like and the intersection
manifold with these planes consists of two ellipses located in the diagonals
of the planes.

\begin{figure}[htbp]
  \begin{center}
    \centerline{\epsfig{file=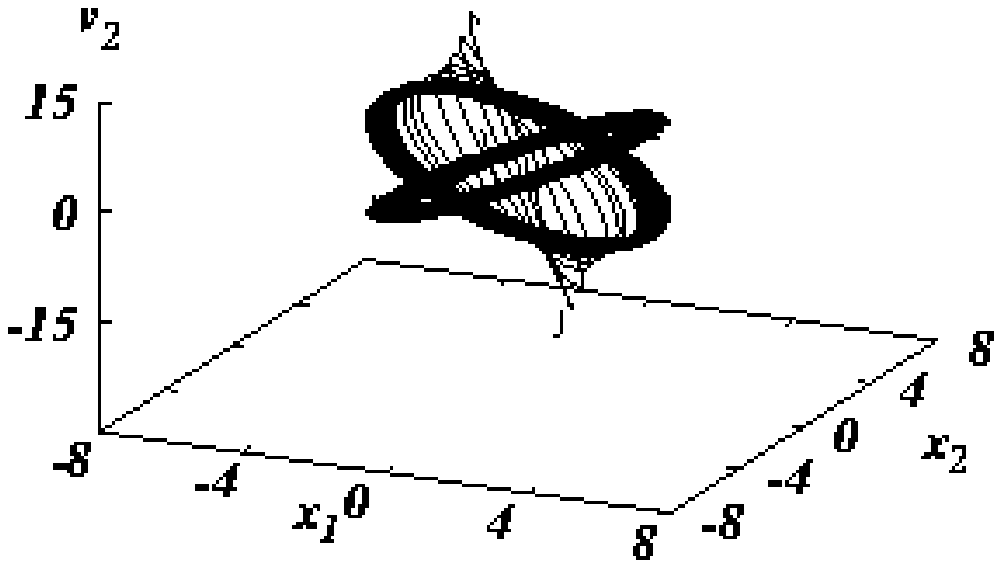,bbllx=126,bblly=301,
        bburx=470,bbury=540,width=8cm}}
    \centerline{\epsfig{file=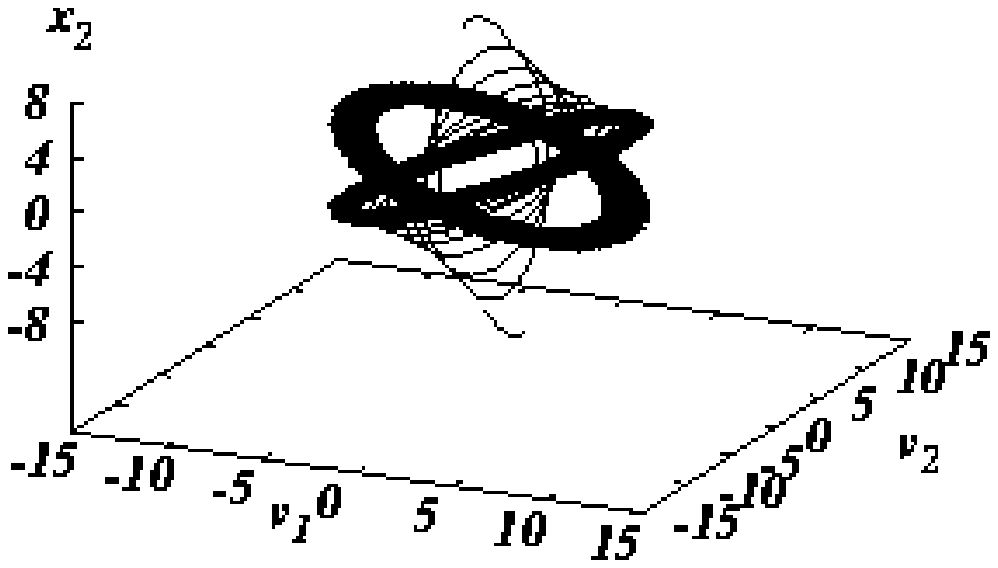,bbllx=126,bblly=301,
        bburx=470,bbury=540,width=8cm}}
    \caption{Projections of a stochastic trajectory on the 
      three-dimensional subspaces $\{x_1,x_2,v_2\}$ and $\{x_1,x_2,v_2\}$ for
      very low noise ($D=0.01$). The particle will not leave the attractor
      which was reached dependant on the initial conditions (One of the
      trajectories represents righthanded and the other on lefthanded
      rotations). The trajectories were obtained with long time computer
      simulations of \eqn{langeq-depot} with constant energy depot, \eqn{e-0}.
      Other parameters: $\gamma_0=0.2$, $c=0.01$, $d_2=0.1$, $a=2.0$ and
      $q_0=10.0$.
      \label{fig:limit-cyc}}
  \end{center}
\end{figure}

\begin{figure}[htbp]
  \begin{center}
    \centerline{\epsfig{file=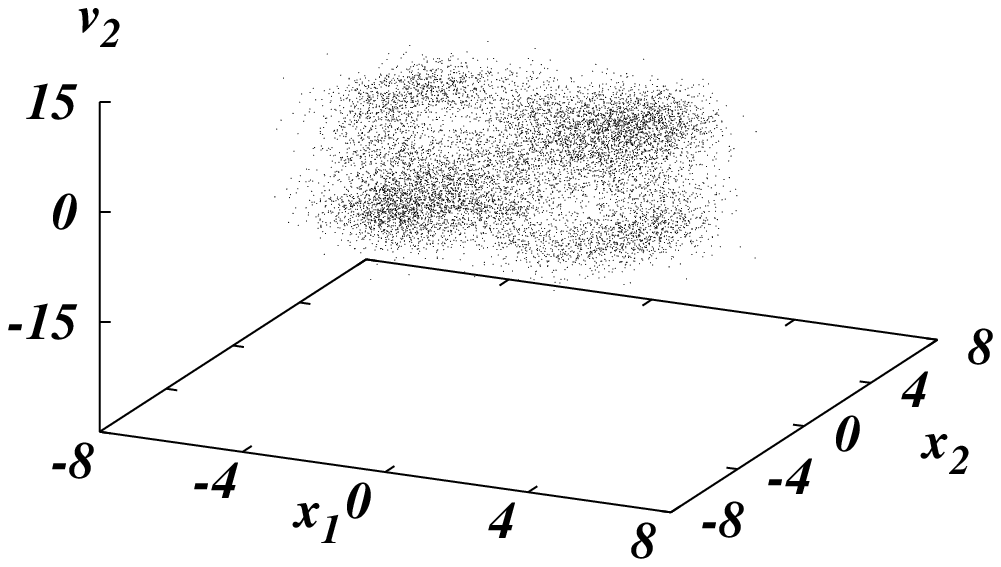,width=8cm}}
    \centerline{\epsfig{file=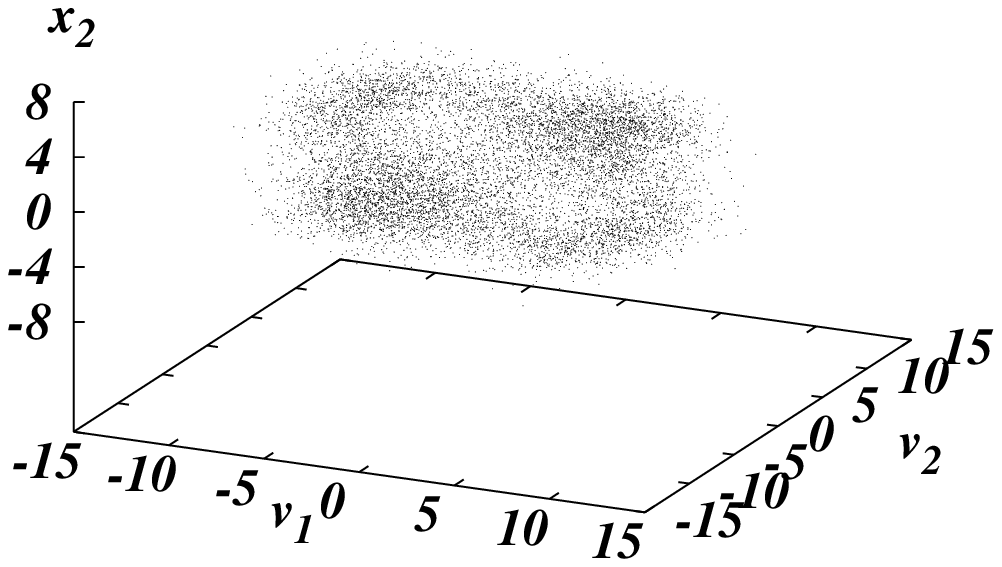,width=8cm}}
    \caption{Projections of a stochastic trajectory on the 
      three-dimensional subspaces $\{x_1,x_2,v_2\}$ and $\{x_1,x_2,v_2\}$ for
      strong noise ($D=0.8$). Other parameters see \pic{fig:limit-cyc}.
      \label{fig:strongnoise}}
  \end{center}
\end{figure}

\begin{figure}[htbp]
  \begin{center}
    \centerline{\epsfig{file=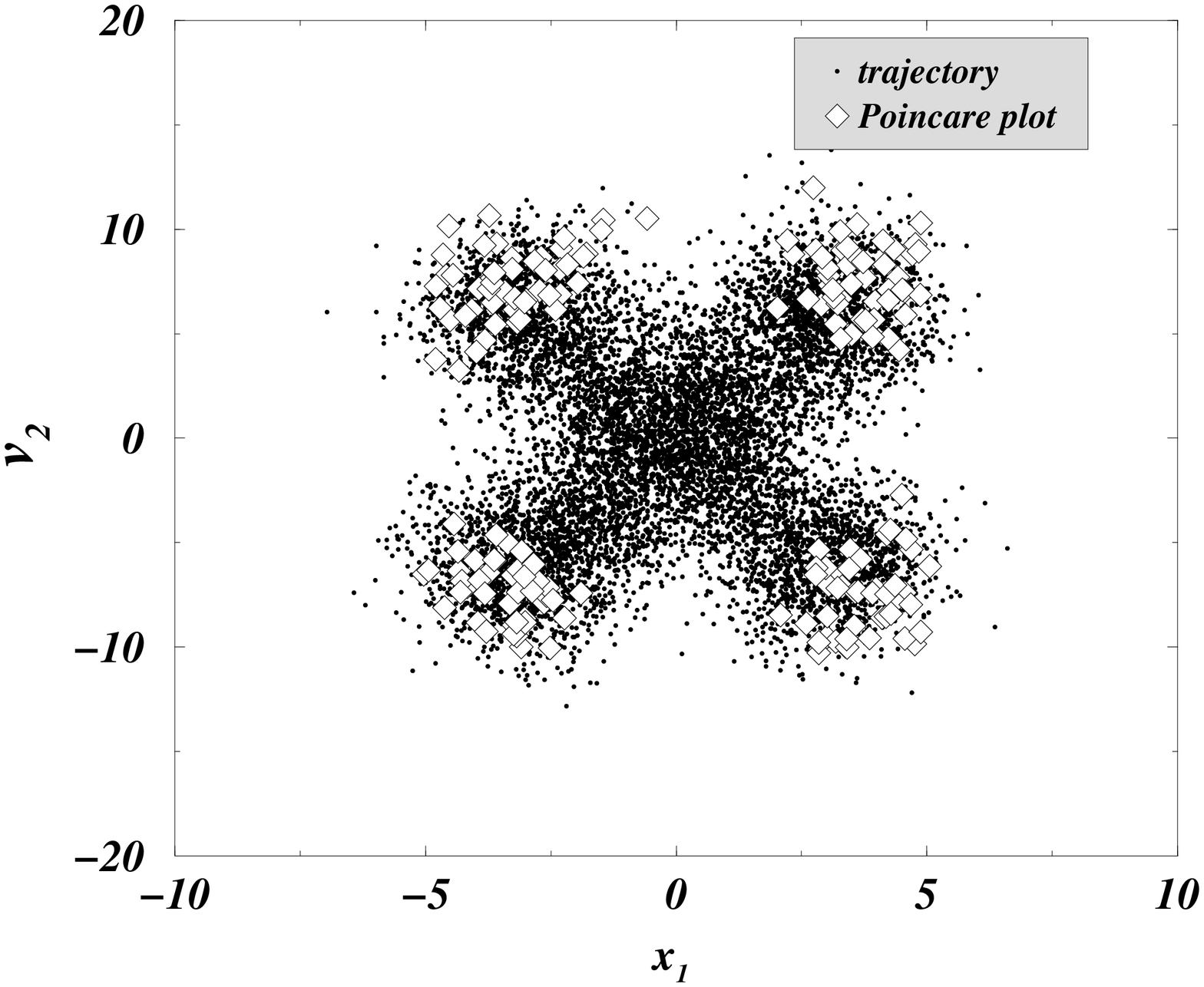,angle=-90,width=7.5cm}}
    \centerline{\epsfig{file=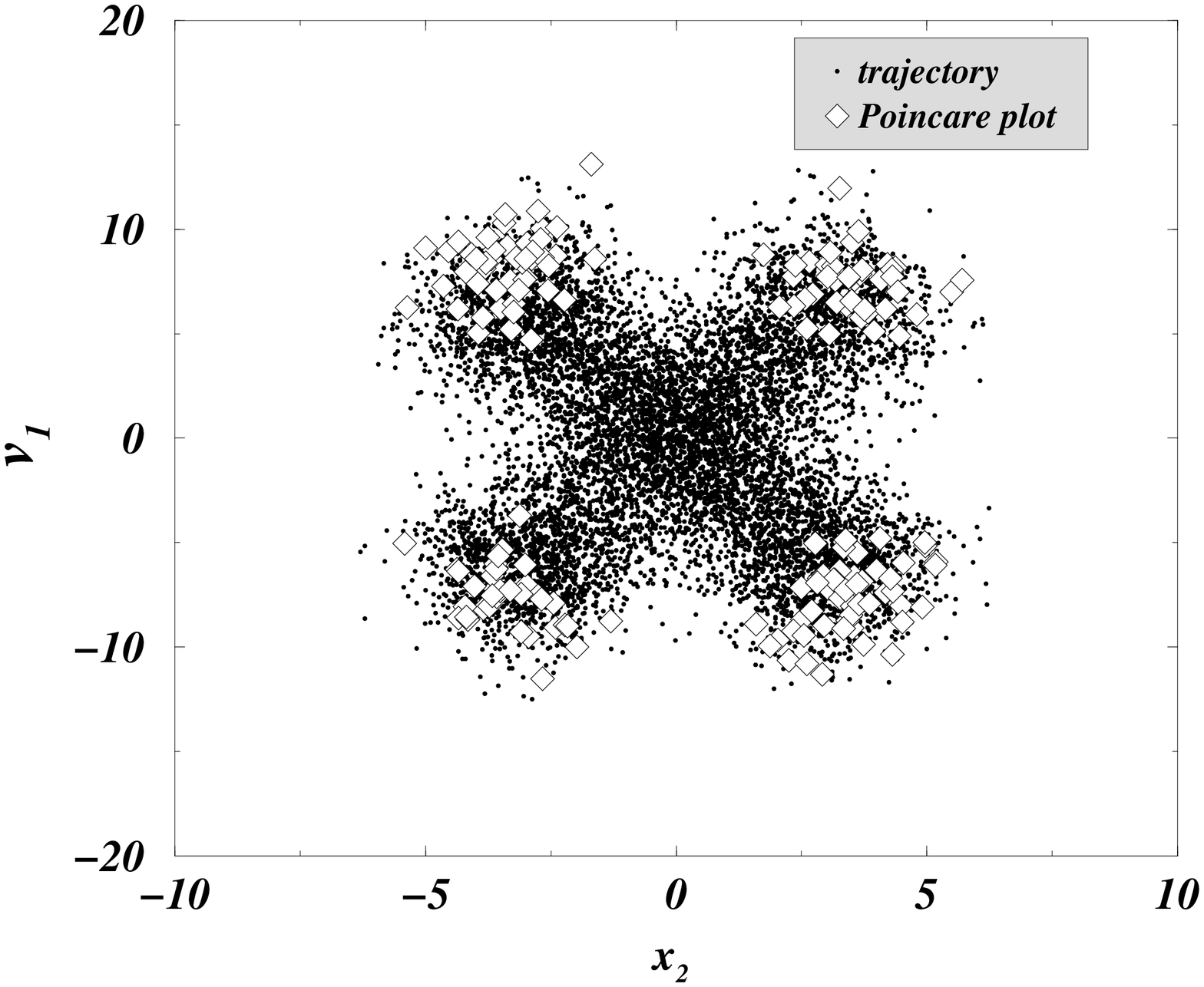,angle=-90,width=7.5cm}}
    \caption{Poincar{\'e} plots and projections of a stochastic trajectory 
      on the two-dimensional subspaces $\{x_1,v_2\}$ and $\{x_2,v_1\}$ (same
      simulation as in \pic{fig:limit-cyc}). The noise strength is large
      enough to allow one particle to penetrate the separatrix between the two
      hula-hoop-like attractors in finite time.  For smaller noise one would
      need to simulate the particle's motion with different initial conditions
      to reach both hoops/tires.
    \label{fig:poincare}}
  \end{center}
\end{figure}

In order to get the explicite form of the distribution we may introduce the
amplitude-phase representation
\begin{equation}
  \label{4d-stochast}
  \begin{array}{rcl}
    x_1 = \rho\, \cos(\omega t + \phi) &\qquad&
    v_1 = - \rho\, \omega\, \sin(\omega t + \phi) \\
    x_2 = \rho\, \sin(\omega t + \phi) &\qquad&
    v_2 = \rho\, \omega\, \cos(\omega t + \phi)
  \end{array}
\end{equation}
where the radius $\rho$ is now a slow \emph{stochastic variable} and the phase
$\phi$ is a fast \emph{stochastic variable}.  By using the standard procedure
of averaging with respect to the fast phases we get for the Rayleigh model of
pumping, \eqn{gamma-ray}, the following distribution of the radii:
\begin{eqnarray}
  P_0(\rho) & \simeq & \exp{\left[2\,\frac {\gamma_1}D a
      \rho^2\left(1-\frac{\rho^2}{r_0^2}\right)\right]} \\
  \nonumber r_0^2 &=& \frac{v_0^2}{\omega^2} = \frac{\gamma_1}{a \gamma_2}
\end{eqnarray}
This distribution has its maximum value at $\rho=r_0$.

For the friction function \eqn{gamma-v2} resulting from the depot model, we
get the following stationary solution for the distribution of the radii:
\begin{equation}
  P_0(\rho) \simeq \left(1 + \frac{d_2}{c} a \rho^2 \right)^{\frac{q_0}{2D}}
  \exp{\left(-\frac {\gamma_0}D a \rho^2 \right)}
\end{equation}
Again, we find that the maximum is located at:
\begin{equation}
  \rho^2=r_0^2=\frac{v_0^2}{\omega^2}=\left(\frac{q_0}{\gamma_0}
    -\frac{c}{d_2}\right)\frac{1}{\omega^2}
\end{equation}
Due to the special form of the attractor which consists, as pointed out above,
of two hula hoops embracing each other, the distribution in the phase space
cannot be constant at $\rho={\rm const.}$ An analytical expression for the
distribution in the four-dimensional phase space is not yet available. But it
is expected that the probability density is concentrated around the two
deterministic limit cycles.

\subsection{Generalization for Other Potentials}
For the harmonic potential discussed above the equal distribution between
potential and kinetic energy, $mv_0^2 = a r_0^2$, is valid, which leads to the
relation: $\omega_0 = v_0/r_0 = \omega$. For other radially symmetric
potentials $U(r)$, this relation has to be replaced by the condition that on
the limit cycle the attracting radial forces are in equilibrium with the
centrifugal forces. This condition leads to
\begin{equation}
  \label{centrif}
  \frac{v_0^2}{r_0} = |U'(r_0)| 
\end{equation} 
For a given $v_0$, this equation defines an implicit relation for the the
equilibrium radius $r_{0}$, namely: $v_0^2 = r_0\, |U'(r_0)|$.  Then the
frequency of the limit cycle oscillations is given by
\begin{equation}
  \label{omega-2}
  \omega_0^2 = \frac{v_0^2}{r_0^2} = \frac{|U'(r_0)|}{r_0} 
\end{equation} 
In the case of linear oscillators this leads to $\omega_0 = \sqrt{a}$ as
before, \eqn{a-12}. But e.g. for the case of a quartic potential
\begin{equation}
  \label{quart}
  U(r) = \frac{k}{4} r^4  
\end{equation} 
we get the limit cycle frequency
\begin{equation}
  \label{omega-4}
  \omega_0 = \frac{k^{1/4}}{v_0^{1/2}} 
\end{equation} 
If \eqn{centrif} has several solutions for the equilibrium radius $r_{0}$, the
dynamics might be much more complicated, e.g. we could find Kepler-like orbits
oscillating between the solutions for $r_0$. In other words, we then find --
in addition to the driven rotations already mentioned -- also driven
oscillations between the multiple solutions of \eqn{centrif}.

In the general case of potentials $U\left(x_1, x_2\right)$ which do not obey a
radial symmetry, the local curvature of potential levels $U\left(x_1,
  x_2\right)= {\rm const.}$ replaces the role of the radius $r_{0}$. We thus
define $\rho_U\left(x_1, x_2\right)$ as the radius of the local curvature. In
general, the global dynamics will be very complicated, however the local
dynamics can be described as follows: Most particles will move with constant
kinetic energy
\begin{equation}
  \label{eq:E=const}
  \bbox{v}^2=v_1^2 + v_2^2 = v_0^2 = {\rm const.}
\end{equation}
along the equipotential lines, $U\left(x_1, x_2\right)={\rm const.}$ In
dependence on the initial conditions two nearly equally probable directions of
the trajectories, a counter-clockwise and a clockwise direction, are possible.
Among the different possible trajectories the most stable will be the one
which fulfills the condition that the potential forces and centrifugal forces
are in equilibrium:
\begin{equation}
  \label{eq:**}
  v_0^2 = \rho_U\left(x_1, x_2\right)\abs{\nabla U\left(x_1, x_2\right)}
\end{equation}
This equation is a generalization of \eqn{centrif}.

\section{Discussion and Conclusion}
\label{sec:4}
An interesting application of the theoretical results given above, is the
following: Let us imagine a system of active Brownian particles which are
pairwise bound by a potential $U(r_1 - r_2)$ to dumb-bell-like configurations.
Thus, the pairs of active particles could be regarded as ``active Brownian
molecules''.  Then the motion of each ``molecule'' consists of two independent
parts: the free motion of the center of mass and the relative motion under the
influence of the potential.

The motion of the center of mass is described by the equations given in Sect.
2, while the relative motion is described by the equations given in Sect. 3.
As a consequence, the center of mass of the dumb-bell will perform a driven
Brownian motion, but in addition the dumb-bell is driven to rotate around the
center of mass. We then observe that a system of pumped Brownian molecules --
with respect to their center of mass velocities -- has a distribution
corresponding to \eqn{RayHelm_stat} or \eqn{p0-v}. However, since the internal
degrees of freedom are excited, we also observe driven rotations and in
general (if \eqn{centrif} has at least two solutions) also driven
oscillations. Thus, we conclude that the mechanisms described here may be used
also to excite the internal degrees of freedom of Brownian molecules.

Another possible application is the motion of clusters of active Brownian
molecules.  Similar to the case of the dumb-bells, these clusters are driven
by the take-up of energy to perform \emph{spontaneous rotations}. Eventually,
a stationary state will be reached which is a mixture of rotating clusters or
droplets, respectively.

These suggested applications for the dynamics of pumped Brownian particles
will of course need further investigations. In this paper the aim was rather
to study the influence of non-linear friction terms on Brownian particles from
a more general perspective.

A suitable expression for a non-linear, velocity-dependent friction function
$\gamma(\bbox{v})$ has been obtained by considering an internal energy depot
of the Brownian particles. This energy depot can be changed by energy take-up,
internal dissipation and conversion of internal into kinetic energy. Provided
a fast relaxation, the depot can be described by a quasistationary value. We
then found that the velocity-dependent friction function describes the pumping
of energy in the range of small velocities, while in the range of large
velocities the known limit of (dissipative) friction is reached.

In this paper, the focus was on the distribution function of pumped Brownian
particles. Therefore, in addition to the mechanisms of pumping, we had to
consider the dependence on stochastic influences, namely on the strength of
the stochastic force, $D$. We could find new dynamical features for pumped
Brownian particles, such as:
\begin{enumerate}
\item new diffusive properties with large mean squared displacement, which
  increases quadratically with the energy supply and scales with the noise
  intensity as $1/D$
\item non-equilibrium velocity distributions with crater-like shape,
\item the formation of limit cycles corresponding to left/righthand rotations,
\item the switch between these opposite directions, i.e. the penetration of
  the separation, during the stochastic motion of a pumped Brownian particle.
\end{enumerate} 
Qualitative similar features are obtained for the motion of driven particles
in physico-chemical systems, for example for the motion of liquid droplets on
hot surfaces, or the motion of surface-active solid particles
\cite{DeDu71,DuDe74,Ge85,MiMe97}. There are also relations to active
biological motion \cite{SchiGr93,EbSchwTi99}. In this paper, we did not intend
to model any particular object but analyzed the general physical
non-equilibrium properties of such systems.  While our investigations are
based on rather simple physical assumptions on non-linear friction for
Brownian particles, we found a rich dynamics, which might be of interest for
more applied investigations later.


\end{document}